\documentstyle[twocolumn,aps]{revtex}

\begin{document}

\title{Wavefunction Intensity Statistics from Unstable Periodic Orbits}
\author{L. Kaplan \thanks{kaplan@physics.harvard.edu}
\\Department of Physics and Society of Fellows,\\ Harvard
University, Cambridge, MA 02138}
\maketitle

\begin{abstract}
We examine the effect of short unstable periodic orbits on wavefunction
statistics in a classically chaotic system, and find that the tail
of the wavefunction intensity distribution in phase space is dominated by
scarring associated with
the least unstable periodic orbits. In an ensemble average over systems
with classical orbits of different instabilities, a power law tail
is found, in sharp contrast to the exponential prediction of random
matrix theory. The calculations are compared with numerical data
and quantitative agreement is obtained.
\end{abstract}

\vskip 0.2in

Quantum eigenstates of classically chaotic systems generically
exhibit a phenomenon known as scarring, the enhancement of intensity
along short unstable periodic orbits for some fraction of the
wavefunctions. Scarring is a fascinating example of the influence of
identifiable classical structures on stationary quantum properties and
on long-time quantum transport in a classically ergodic system. The
occurrence of scars is in some sense paradoxical, because classically,
all such short-time information is destroyed at long times, and a
classical probability distribution after being evolved for a sufficiently
long time retains no memory of its initial state. Scarring is one
of the most dramatic examples of a departure of quantum chaotic
systems from the predictions of random matrix theory (RMT), according
to which wavefunctions
must be evenly distributed over phase space, up
to quantum fluctuations. Scarring has now been observed experimentally
in a variety of systems, including microwave cavities\cite{sridhar,stockman},
tunnel junctions\cite{fromhold},
and the hydrogen atom in a uniform magnetic field\cite{wintgen,wintscar}.

Examples of scarring were observed numerically in \cite{hellerscar},
and a theory
based on the semiclassical evolution  of Husimi states near
a periodic orbit was provided. Later work by Bogomolny \cite{bogomolny}
and Berry \cite{berry} involved calculations in coordinate space and
Wigner phase space, respectively.
All these works were based on
the linearized dynamics around the unstable periodic orbit, and were
thus, by construction, theories of the short-time behavior only.
Yet to get a true understanding of the properties of individual eigenstates
it is essential to understand the long-time quantum dynamics,
including returns of amplitude to the original periodic orbit after
undergoing excursions into other areas of phase space.
In a recent paper \cite{nlscar}, a formalism was developed for dealing with
these nonlinear contributions to scarring, providing quantitative agreement
of the theory with numerical results. This work used a measure of
scarring based on Husimi intensities. (Recently Fishman, Agam, and others
have provided interesting new perspectives on the problem on scarring, and have
offered a measure of scarring related to, but somewhat different from ours
\cite{fishmanagametc}. A number of other authors have also made significant
contributions in this area; we cannot list them all but
a few recent references are provided in
\cite{others}.)
In this paper, we will
apply a result previously obtained in \cite{nlscar}
to compute the effect of scarring on 
the wavefunction intensity distribution, a function
which has been investigated previously for diffusive systems\cite{diffuse}.

We first give a general idea of the formalism and state the key result
of \cite{nlscar} which is relevant
to the present work. A gaussian wavepacket $|\Psi\rangle$ is initially centered
close to a classical periodic orbit and allowed to evolve. If the
instability exponent $\lambda$ of the orbit is small, the 
function $A(t)=\langle\Psi|\Psi(t)\rangle$
will (under the linearized dynamics) exhibit recurrences on a time
scale of one period $T_P$, and the
amplitude of these recurrences decays on a scale $T_P/\lambda$. This decay
leads to the formation of envelopes in the local density of states
$S(E)$, which
is the fourier transform of the autocorrelation function. These envelopes
in the energy spectrum have spacing $\hbar/T_P$, width scaling as
$\hbar \lambda/T_P$ for small $\lambda$, and height scaling as
$\lambda^{-1}$. Long-time (nonlinear)
recurrences lead to fluctuations multiplying
this envelope in the energy domain. Eventually, by the Heisenberg time
$T_H=\hbar/\Delta$, where $\Delta$ is the mean level spacing, individual
peaks are resolved in the spectrum $S(E)$, the heights of the peaks
being the intensities of the corresponding eigenstates at the test state
$|\Psi\rangle$, $S(E)=\sum_n |\langle n|\Psi\rangle|^2 \delta(E-E_n)$.

The simplest picture of the non-linear recurrences assumes that the
periods, actions, and homoclinic points corresponding to the long-time
excursions are random and uncorrelated, up to the constraint of
unitarity which produces delta-function spectral peaks by the
Heisenberg time. This, when combined with the known short-time
dynamics in the linear regime, can be shown to produce a chi-squared
distribution of spectral intensities multiplying the original linear
envelope.\footnote{
Strong, isolated nonlinear recurrences which cannot be treated statistically,
as well as correlations between long-time excursions modify this simple
picture. However, we will find that for the generalized baker's maps (a
paradigmatic example of hard chaos), the assumptions of randomness beyond
the linear decay time lead
to results which are in quite reasonable agreement with numerical data.}

A key result of the calculations in Ref.~\cite{nlscar} is the following:
{\it individual spectral lines (overlap intensities) in the local density of
states obey the usual chi-squared (Porter-Thomas) fluctuations, but these
are modulated by the fourier transform of the linearized short-time
autocorrelation function.}
The latter
can be computed analytically in terms of the instability $\lambda$
of the classical orbit in question (see Eq.~\ref{alin} below).
More explicitly, in the case of complex eigenstates,
the chi-squared distribution has two degrees of freedom, and in the absence
of scarring the probability of having a spectral line height greater
than $x$ is given by $P(x)=\exp(-x)$. Here $x$ is normalized to have a mean
value of unity, i.e. $x_n=N |\langle n|\Psi\rangle|^2$, where $N$ is the total
number of states. Now in the
presence of scarring this is modified to
\begin{equation}
\label{pscar}
P(E,x)=\exp{(-x/S_{\rm lin}(E))}
\end{equation}
for a eigenstate with energy $E$. {\it Here $S_{\rm lin}$
is the spectral envelope
given by the fourier transform of the linearized dynamics} (Eq.~\ref{alin}
for the case treated in the present paper).

In the following discussion we will for simplicity consider the case
of scarring by a fixed point of a discrete-time map. (The generalization
of the results to orbits of period greater than one and to continuous
time systems is straightforward\cite{nlscar}.)
So let us consider without loss
of generality a hyperbolic fixed point at the origin of a compact phase space,
with exponent $\lambda$, and stable and unstable manifolds oriented along
the $p$
and $q$ axes, respectively. The equations of motion near the fixed point are
then given by
\begin{eqnarray}
\label{eqmo}
q' & = & q \exp(\lambda t)  \nonumber \\
p' & = & p \exp(-\lambda t) \,.
\end{eqnarray}
In the presence of shearing, e.g. $\partial q' / \partial p \ne 0$, or for
non-orthogonal stable and unstable manifolds, or in the case where the
manifolds are not oriented along the $p$ and $q$ axes, a canonical
transformation would first need to be performed to get the equations
of motion into the form above. 
Now we define our test state to be a gaussian
wavepacket centered at $(q_0,p_0)$, with horizontal width $\sigma$ and
vertical width $\sigma_p=\hbar/\sigma$. In coordinate representation
this is given by
\begin{equation}
\label{psiform}
\Psi(q)=\left(4\pi \sigma_p^2 \right)^{1/4}
\exp{[-(q-q_0)^2/2\sigma^2+ip_0(q-q_0)/\hbar]} \,.
\end{equation}
In situations described above where the local equations of motion
do not have the form of Eq.~\ref{eqmo} in the natural coordinates,
an optimal test state would have
a complex width $\sigma$ in those coordinates, as can be
seen by performing a canonical transformation of the gaussian
of Eq.~\ref{psiform}.

We now allow the  wavepacket to evolve under the linearized dynamics,
stretching each time step
by a factor $e^\lambda$ in the $q-$direction and shrinking
by the same factor in the $p-$direction. The linearized quantum autocorrelation
function  after time $t$ is given by
\begin{eqnarray}
& & A_{\rm lin}(q,p,\sigma,\lambda,t)={\exp{i\theta_0 t}
  \over \cosh{\lambda t}} \nonumber
\\ & & \times \exp\left(-{\cosh{\lambda t}-1 \over 2 \cosh{\lambda t}}
\left({q_0^2\over\sigma^2}+{p_0^2
\over\sigma_p^2}\right)-{iqp\over \hbar}\tanh{\lambda t}\right) \,.
\label{alin}
\end{eqnarray}
Here $\theta_0$ is the phase associated with one iteration of the
periodic orbit, given by the classical action in units of $\hbar$, plus
any Maslov indices associated with caustics in the classical dynamics.
$\theta_0$ determines the location of the peak in the spectral envelope
$S_{\rm lin}(E)$, defined to be the fourier transform of
$A_{\rm lin}(t)$. Since we will be interested in performing an energy average,
we freely set $\theta_0=0$.

Now the expression in Eq.~\ref{alin} can be inserted into Eq.~\ref{pscar}
to obtain the distribution of wavefunction intensities at a given energy.
Because we wish here to consider all eigenstates, independent of energy,
we then perform an energy averaging,
remembering that for a map the quasi-energy is defined to lie between
$0$ and $2\pi$ only.
We also notice that the tail of
the intensity distribution will be dominated by the peak of
the spectral envelope at $E=0$, and we therefore can use a saddle
point approximation, obtaining
\begin{eqnarray}
P(q,p,\sigma,\lambda,x) & = & {1\over 2 \pi} 
\int dE P(q,p,\sigma,\lambda,E,x) \nonumber \\
& \approx & {1 \over \sqrt{2 \pi}} {\exp(-x/S_{\rm lin}(q,p,\sigma,\lambda,E=0))
\over \sqrt{{-x \over S_{\rm lin}(0)^2}
{\partial^2 S_{\rm lin}\over \partial E^2}}} \,,
\end{eqnarray}
where the expression obtained is an asymptotic form valid for large $x$.
For small $\lambda$, the sum over time steps can be replaced by
an integral, and we have at $q_0=p_0=0$
\begin{equation}
S_{\rm lin}(E)=\int dt {e^{-i Et} \over \sqrt{\cosh{\lambda t}}} \,.
\end{equation}
Now by dimensional analysis, $S_{\rm lin}(0)=Q/\lambda$
and ${\partial^2 S_{\rm lin}\over \partial E^2}(0)=-W/\lambda^3$,
where $Q$ and $W$ are numerical constants. We thus obtain the first
result of this paper, the tail of the intensity distribution
for a wavepacket centered on a periodic orbit,
\begin{equation}
\label{res1}
P(q_0=0,p_0=0,\sigma,\lambda,x) =
{1 \over \sqrt{2 \pi}} {Q \over \sqrt W} \lambda (x \lambda)^{-1/2}
 e^{-x\lambda/Q} \,.
\end{equation}
Notice that the exponential tail
has been effectively stretched by a factor of $Q/\lambda$, corresponding
to the increased height of the peak of the linear envelope at small
$\lambda$. There is also a linear suppression factor of $\lambda$,
corresponding to the width of the peak in $S_{\rm lin}(E)$, and
indicating that only a fraction scaling as $\lambda$ of all the eigenstates
are effectively scarred. The remainder of the eigenstates are
typically ``antiscarred", having on average a lower intensity at the periodic
orbit than would be expected based on RMT. This distribution will
have a nontrivial inverse participation ratio (IPR), scaling as the
inverse of the width of the linear envelope, {\it i.e.} as $1/\lambda$.

The region of validity of Eq.~\ref{res1} is 
\begin{equation}
1 \ll \lambda^{-1} \ll x \ll N \,.
\end{equation}
The
first inequality ensures that many iterations of the periodic orbit
contribute (so the sum over iterations can be replaced by
an integral) and the scarring is strong. In fact, however, because of
the large value of the numerical constant $Q$, the formula works well
even for exponents as small as $\log 2$, as will be seen in the
numerical study below. The second inequality  says that we are in the
tail of the distribution and the events are all coming from the
peak of the linear envelope. The third inequality is a unitarity
constraint -- obviously our assumption of random fluctuations breaks
down for intensities of order unity, when the entire wavefunction
would be concentrated in a phase space area of order $\hbar$.

Now we go on to perform a similar analysis integrating over the phase
space variables $q_0$ and $p_0$. As before, we take the
exponential $\exp{(-x/S_{\rm lin}(q_0,p_0,\ldots))}$
and expand to second order in $q_0,p_0$
around the maximum $q_0=p_0=0$.
Then upon integration by stationary phase we pick up a determinant factor of
\begin{equation}
{2\pi S_{\rm lin}(0)^2 \over x} {1\over\sqrt{
{\partial^2 S_{\rm lin}\over \partial q_0^2}{\partial^2
S_{\rm lin}\over \partial p_0^2}}}\,.
\end{equation}
Here we have taken the classical phase space volume to be unity
for simplicity. 
Now for small $\lambda$,
\begin{equation}
\label{pswidth}
{\partial^2 S_{\rm lin}\over \partial q_0^2}(0)=
{2\over \sigma^2} {-Z\over\lambda} \,,
\end{equation}
where $Z$ is yet another numerical constant,
and similarly for $p_0$, with $\sigma$ replaced by $\sigma_p$=$\hbar/\sigma$.
So the total factor resulting from the phase space integration is
${\pi \hbar\over (x\lambda)} {Q^2\over Z}$, again
independent of
$\sigma$. Combining this with the
expression in Eq.~\ref{res1}
above, we obtain the second result, for the distribution
of overlap intensities after energy and phase space averaging,
\begin{equation}
\label{res2}
P(\lambda,x)=
\sqrt{\pi \over 2} {Q^3 \over Z \sqrt W} \hbar \lambda
(x\lambda)^{-3/2}\exp{(-x\lambda/Q)} \,.
\end{equation}
Here we have picked up a factor of $\hbar$ from the factor of $\sigma$
in Eq.~\ref{pswidth} and the corresponding factor of $\sigma_p=\hbar/\sigma$
associated with the falloff in $S_{\rm lin}$ in the momentum direction.
This indicates that the tail
is coming entirely from the region near the periodic orbit, specifically
from wavepackets that have large classical probability density right on
the orbit.  (Thus,
a measure like the IPR for a generically placed wavepacket
will not see the effect of scarring by an individual
periodic orbit, when the semiclassical limit has been taken.)
The result in Eq.~\ref{res2} is valid in the regime
\begin{equation}
\max(\log N,\lambda^{-1}) \ll x \ll N \,.
\end{equation}
Here $\log N$ is the value of $x$ near which the RMT exponential
decay law reaches values of order $\hbar=1/2\pi N$.
In this region, a crossover occurs between the head
of the distribution, which is dominated by non-scarred region
of phase space and approaches the Porter-Thomas (RMT)  prediction, 
and the tail, dominated by scarring,  given by the expression above.
The  expression Eq.~\ref{res2} holds also for an ensemble of
systems, all having  an orbit with instability $\lambda$. In principle,
we should of course do a sum over all periodic orbits, 
however the tail will clearly
always be dominated by the orbit with smallest $\lambda$.

Finally, we now consider an ensemble of systems
where the value of the smallest
exponent varies from system to system, with distribution
${\cal P}(\lambda)=C \lambda^\alpha$ for small $\lambda$.
Then using Eq.~\ref{res2} and integrating over  $\lambda$ we obtain
\begin{equation}
\label{res3}
P(x)=C \sqrt{\pi \over 2} {Q^3 \over Z \sqrt W}
Q^{\alpha+1/2}\Gamma(\alpha+1/2)
\hbar x^{-(2+\alpha)} \,.
\end{equation}
Note that this is an uncontrolled approximation because
we have integrated over $\lambda$ after having assumed
$x \lambda$ was large. However, if we had included higher-order
corrections in $(x \lambda)^{-1}$ in Eq.~\ref{res2}, the scaling of
$P(x)$ would remain unchanged, i.e.
\begin{equation}
P(x)= C f(\alpha) \hbar x^{-(2+\alpha)} \,,
\end{equation}
with the dimensionless function $f(\alpha)$ somewhat different
from that given in Eq.~\ref{res3}. An important point is that the
tail displays power-law behavior in the intensity $x$, a strong
deviation from the exponential prediction of RMT. As with Eq.~\ref{res2},
this asymptotic form is valid for values of $x$ large compared to
$\log N$ and small compared
to $N$. For small $x$ we again
expect a crossover to the Porter-Thomas form. For large $x$
we expect a downward correction away from  the $x^{-(2+\alpha)}$ form,
with a breakdown of the approximation occurring at some
fraction of $N$, depending on $\alpha$.

Now, we proceed to test numerically these predictions of the
nonlinear scarring theory. The system we use for this purpose is the
generalized three-strip baker's map, described in some detail
in \cite{nlscar}. This system is similar to the original baker's map,
except that the two strips are replaced by three,
with widths generally unequal, but normalized to $\sum_{i=0}^2 w_i=1$.
There is a fixed point of the classical
dynamics associated with the middle strip, and the instability 
associated with this orbit is given by $\lambda=|\log w_1|$. So we choose
$w_1=1/2$, set $N=200$, and find numerically the wavefunction intensity
distribution at the fixed point
after ensemble averaging over the widths $w_{0,2}$.
(The predictions are
expected to hold for individual systems as well, at sufficiently
large values of $N$. However, for the matrices which we can efficiently
diagonalize, $N$ is not large enough to obtain good statistics in the
tail without ensemble averaging.) A circular
wavepacket of width $\sigma=\sqrt \hbar$ is used.
The results are compared in
Fig.~\ref{fig1} (upper thick curve)
with the prediction of Eq.~\ref{res1}, plotted as a
dashed curve. The RMT prediction is shown as a dotted line.
Note that the theoretical prediction of Eq.~\ref{res1} has no free
parameters and is $\hbar$-independent, depending only on the exponent
$\lambda$ of the periodic orbit in question.

Next, we perform a phase space average for the systems described above,
collecting statistics for wavepackets uniformly distributed over
the entire phase space. The resulting statistics are also plotted
in Fig.~\ref{fig1} (lower thick curve),
where the theoretical prediction for the tail,
given by  Eq.~\ref{res2}, is shown as a solid curve. Again, the 
Porter-Thomas distribution appears as a dotted line. We see a crossover
between the two regimes at a value of  $x$ of order $\log N$.

Finally, we want to construct an ensemble which will contain systems
with orbits of different instability exponents $\lambda$. For this
purpose, we take a uniform distribution of strip widths $w_0$ and $w_2$,
each in the range $[0,1/4]$. The fixed point in the middle strip,
with exponent $\lambda=|\log(1-w_0-w_2)|$, is always the least
unstable periodic orbit. This ensemble has power $\alpha=1$ and
$C=4^2=16$ in the notation of Eq.~\ref{res3}. Averaging
over $100$ systems, we obtain the statistics plotted in Fig.~\ref{fig3}.
The power-law prediction given by  Eq.~\ref{res3} is plotted as a solid
line on the log-log scale, with the RMT prediction as a dotted curve.
Once again, we see a crossover between the two regimes
for $x$ of order $\log N \approx 5.3$. We also see a gradual breakdown
of the approximation as $x$ approaches values comparable to $N=200$. Note
that the quantitative agreement is in spite of the fact that
an uncontrolled  stationary phase approximation was used in obtaining
the overall constant in front of Eq.~\ref{res3}, as explained above.
The important thing to notice here is the power-law behavior of the tail,
in agreement with the theory, and the dramatic deviation from the
predictions of RMT. By $x=100$, where the approximation $x \ll N$ is clearly
beginning to break down, the measured probability is still within a
factor of $4$ of our prediction and is enhanced by $10^{37}$
over the Porter-Thomas value.

We have also checked the linear $\hbar$-dependance of the phase-space
averaged results Eqs.~\ref{res2},\ref{res3} by repeating the preceding
numerical analysis with larger matrices ($N=500,1000$).
In addition, we have constructed
an $\alpha=0$ ensemble by imposing the restriction $w_0=w_2$ and have
observed a $x^{-2}$ power-law behavior in accordance with Eq.~\ref{res3}.

\section{Acknowledgements}

This research was supported by the National Science Foundation under
grant number CHE-9321260. It is a pleasure to thank E. J. Heller for
many stimulating and useful discussions.

\begin{figure}
\caption{Cumulative
wavefunction intensity distribution (a) as measured by a test state
centered on a periodic orbit with instability $\lambda=\log 2$, plotted as
the upper thick curve with
scarring theory prediction given by dashed curve, and (b) averaged over the
entire phase space of size $200h$, plotted as lower thick curve
with theory given by solid curve. The dotted
line is the Porter-Thomas law.}
\label{fig1}
\end{figure}

\begin{figure}
\caption{Cumulative wavefunction intensity distribution after ensemble
averaging over systems with classical orbits of different
instability exponents.
Here again $N=200$, and the dotted curve is the RMT prediction.}
\label{fig3}
\end{figure}

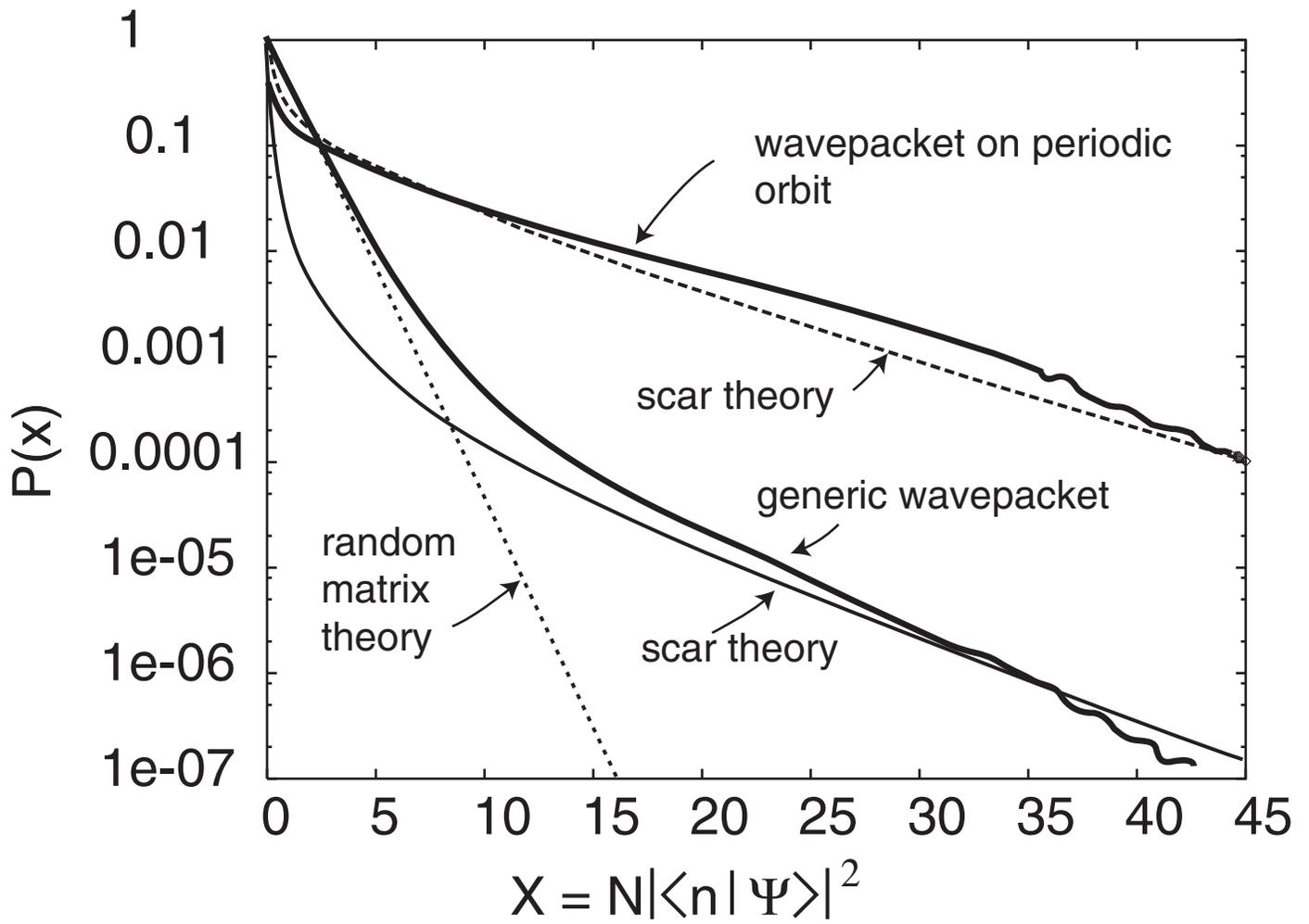
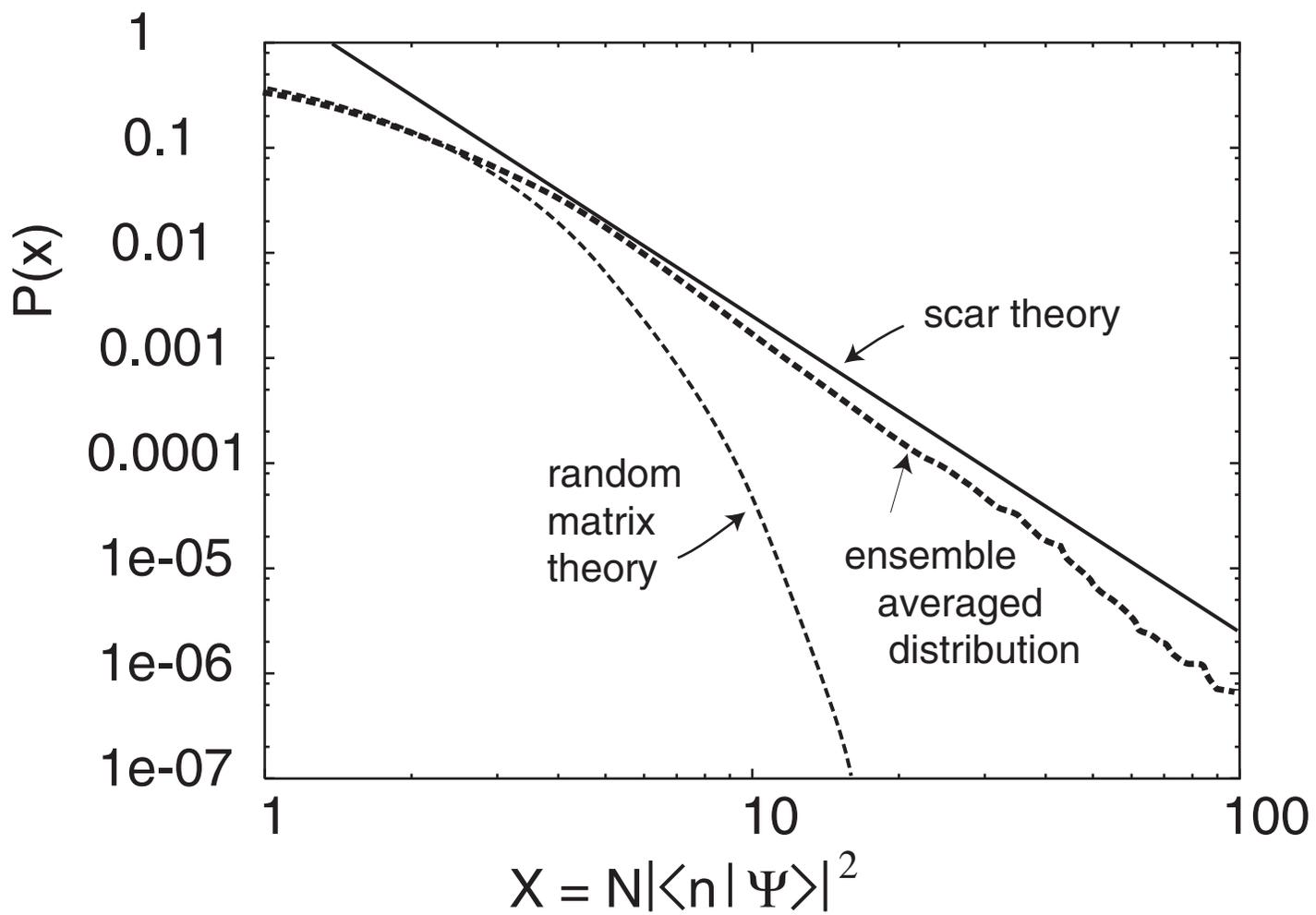
\end{document}